# Energy Efficient Algorithms and Power Consumption Techniques in High Performance Computing

## 8[th] JKSC Srinagar, J&K, India

**Vivek Chalotra[*], Anju Bhasin, Anik Gupta, Sanjeev Singh Sambyal, Sanjay Mahajan**

*Department of Physics, University of Jammu, Jammu 180006, India*

*\*Email ID: vivekathep@gmail.com*

**Abstract:**

High Performance Computing is an internet based computing which makes computer infrastructure and services available to the user for research purpose. However, an important issue which needs to be resolved before High Performance Computing Cluster with large pool of servers gain widespread acceptance is the design of data centers with less energy consumption. It is only possible when servers produce less heat and consume less power. Systems reliability decreases with increase in temperature due to heat generation caused by large power consumption as computing in high temperature is more error-prone. Here in this paper our approach is to design and implement a high performance cluster for high-end research in the High Energy Physics stream. This involves the usage of fine grained power gating technique in microprocessors and energy efficient algorithms that reduce the overall running cost of the data center.

## Introduction

- Conservation of energy has become a major topic nowadays.
- Information & Communication Technology as a whole is estimates to cover 3% of world's $CO_2$ emissions.
- HPC (High-Performance Computing) is no exception: growing demand for higher performance increases total power consumption.
- Efforts made to reduce energy consumption in HPC:-

    - Fine-grained power gating in microprocessors.

    - Energy-efficient hardware

    - Energy efficient algorithms for HPC.

    - Shutting down HW components at low system utilization.

- This work presents energy consumption techniques in designing HPC Cluster for High Energy Physics research.

## What is HPC ?

High-performance computing (HPC) is the use of parallel processing for running advanced application programs efficiently, reliably and quickly. The term applies especially to systems that function above a teraflop or 1012 floating-point operations per second. The term HPC is occasionally used as a synonym for supercomputing, although technically a supercomputer is a system that performs at or near the currently highest operational rate for computers. The most common users of HPC systems are scientific researchers, engineers and academic institutions.

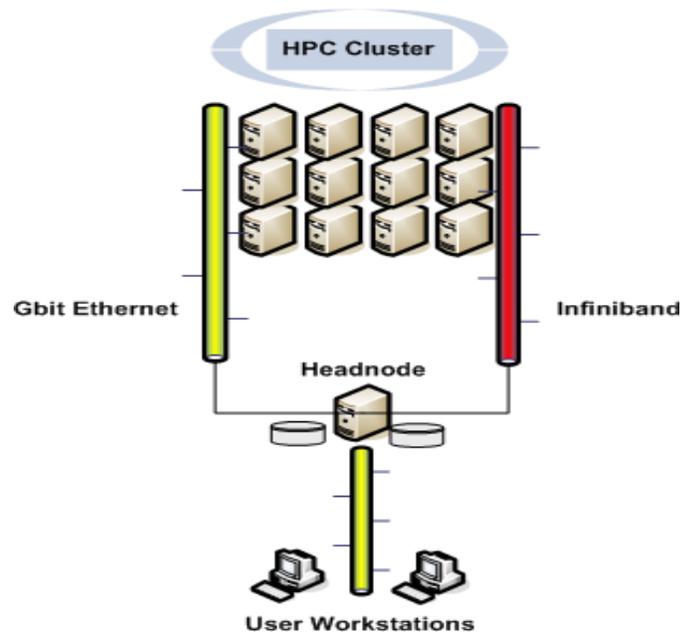

## **Fine-Grained Power Gating**

There are two parts of a processor that consume power: the data-path and memory. The data-path performs computations and takes control decisions, while memory stores data. The waste is built-in. Computing rarely requires everything that a processor is capable of all the time, but all of the processor is fully powered just the same.

One attempt to improve power dissipation in processors is through something called **coarse gating**. It switches off an entire block of the processor that is not being used.

The problem with this method is that most of the time, some part of every component is being used in a processor. Finding an entire block that is not being used at a given time is tough.

**Fine-grained gating** idea is to shut off only the parts of a component that are not being used at the time. While the addition component needs to be capable of adding extremely large numbers, it rarely needs to actually add large numbers. The processor might be using the addition block constantly, but the parts needed to add large numbers can be turned off most of the time. Memory works the same way. A processor needs to be capable of storing large numbers, but seldom actually stores them.

## Power Consumption using Power-Gating

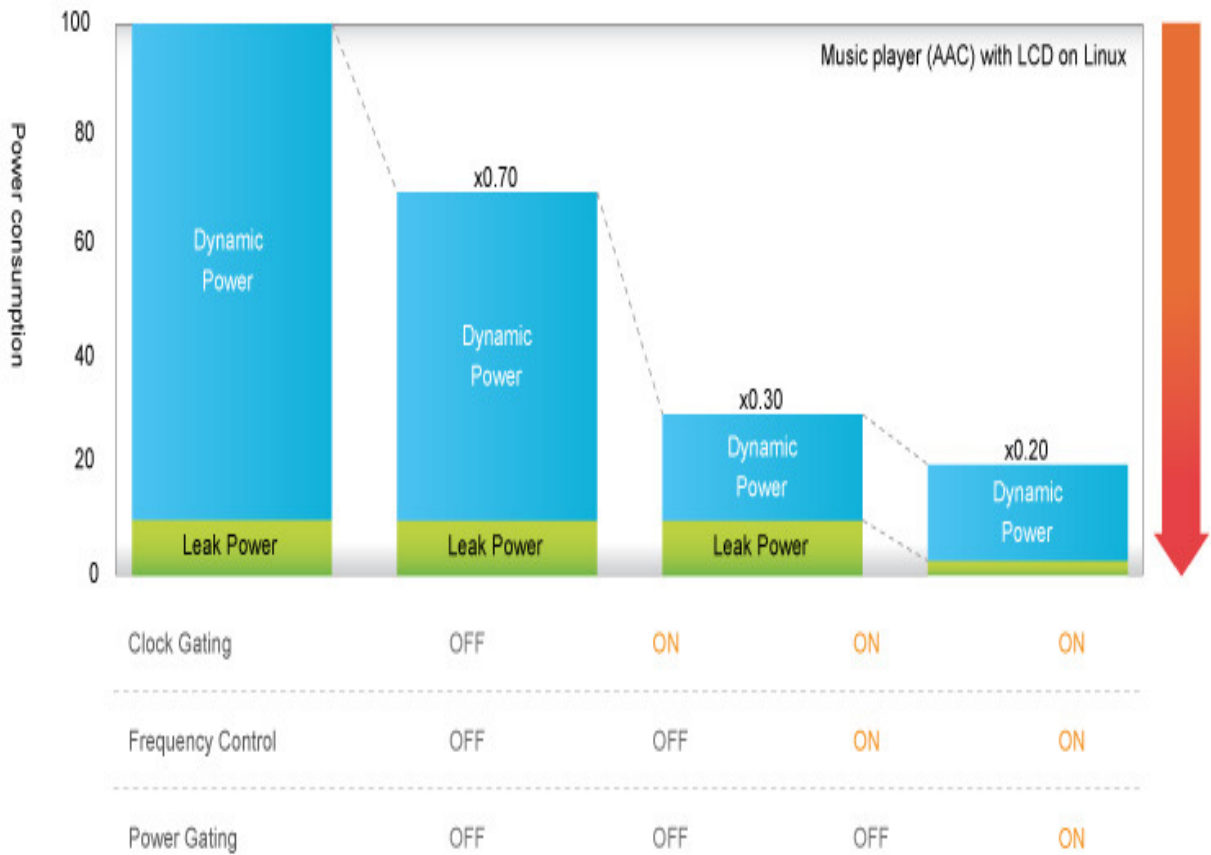

## Energy requirements of Data center

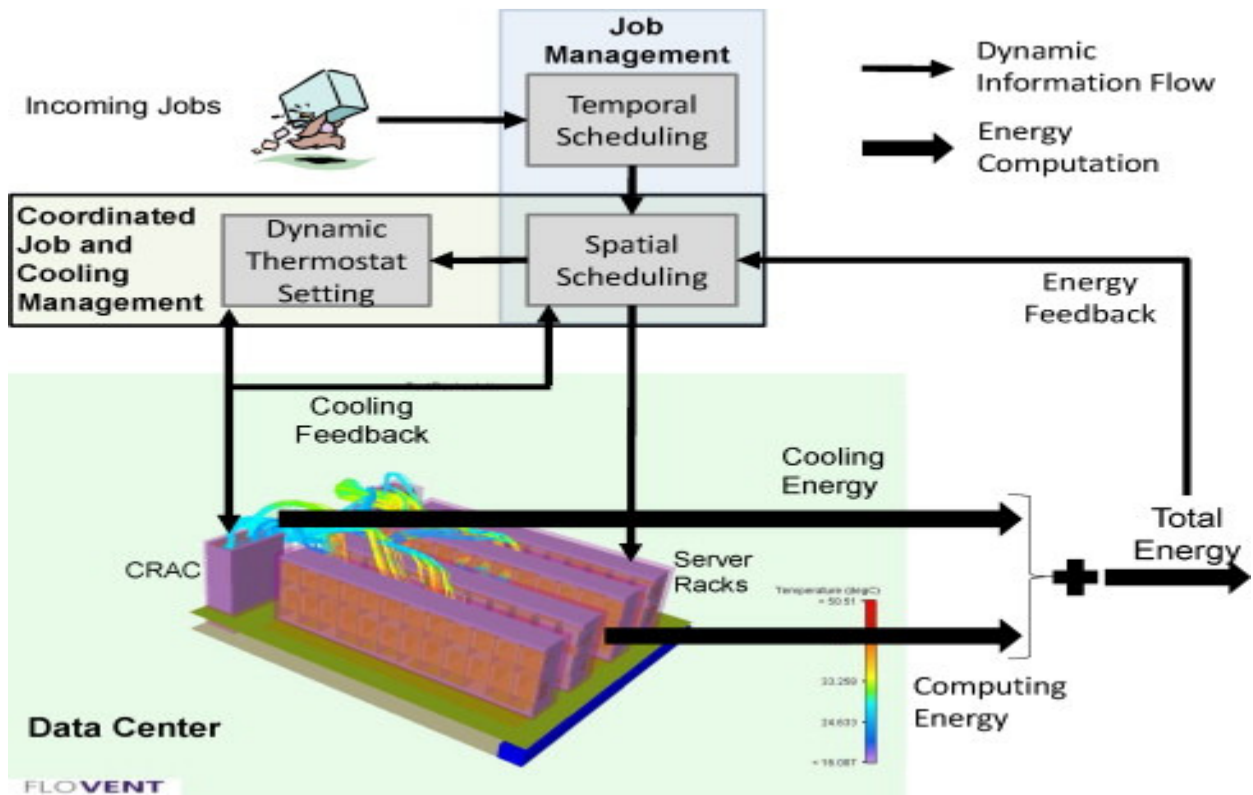

## HGA (HGreen Algorithm)

1. //Statements and Definitions
2. T= {t1, _ _ _ , tn} // Set of workflow tasks
3. R= {r1, _ _ _ , rn} // Set of available grid resources
4. ET← { } // List of tasks ranked by decreasing energy waste
5. ER← { } // List of resources ranked by decreasing energy efficiency
6. pos({L , criterion} // insert position of new element at list L
7. weight {o} // The weight of o object {a task or resource}
8. dil {t} // data intensive level of task t
9. cpe {r} // SPECpower value to r resource
10. ipc {t,r} // Instructions per cycle for task t on resource r
11. iopsw (r) // iops per watts rate for storages
12. gf - Green Factor // 0 , , 1
13. // Analysis Phase
14. for each t in T do
15. weight {t} = analyzer {t}
16. end for
17. // Prioritizing Phase
18. for each t in T do
19. v = pos {ET, weight {t} }
20. Add {t, ET, v} // add t task to ET list at v position
21. end for
22. // Mapping Phase
23. While 3t in ET do
24. t= ET.get{0} // the most power cost task
25. If ready {t} // in case parent tasks of t have been
    completed
26. for each r in R do
27. ee{r} = gf* {cpe{r} + dil {t} * iopsw{r} + ipc{t, r}
28. u = pos{}ER, weight{r} }
29. Add {r, ER, u} // adds resource r to list ER at pos u
30. end for

31. re = ER.get {0} // The most energy efficient resource

32. Schedule {t, re } // assigns t task to the re

33. end if

34. Remove {ET, 0} // removes first item list

35. end while

**Energy-Efficient Grid Architechture**

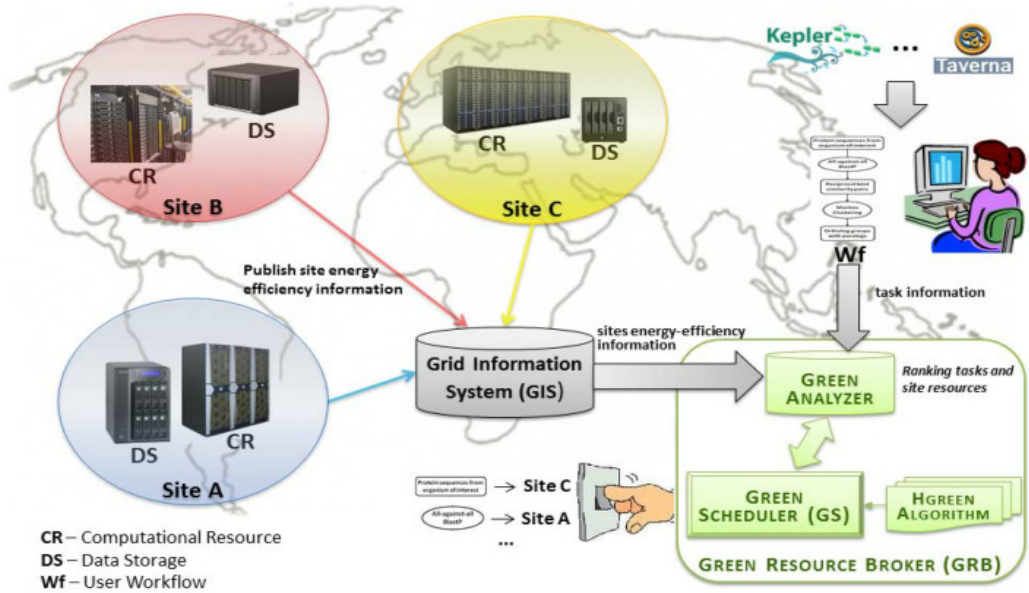

This section describes **Energy-Efficient Grid Architecture (EEGA)** comprises traditional grid components combined with green components in charge of implementing **HGreen Heuristic**. Fig above depicts EEGA having green components inside squared box. Grid sites maintain clusters at different locations all over the world. Fig above shows 3 distinct sites: A, B & C. Each site provides Computational Resource (CR) and Data Storage(DS), responsible for task execution and data management, respectively. These entities play a central role on the Energy Efficient Model, as their energy efficiency is analyzed.

**Summary and Conclusions**

- ✓ We presented a n Energy Efficient Cluster based study.
- ✓ HGA (HGreen Algorithm) can have an effect on energy consumption.
- ✓ Energy savings of 10%-15% can be achieved using this study.
- ✓ Lot of energy can be saved using Fine-grained power gating technique.
- ✓ HEP research needs this kind of HPC.
- ✓ **References**